# Stellar wind variability in OB stars and its effect on mass-loss determinations

**Raman K. Prinja**

Department of Physics and Astronomy, University College London, Gower Street, London WC1E 6BT, UK; r.prinja@ucl.ac.uk

**Abstract**

Virtually all stellar winds of OB stars exhibit variability in one spectral diagnostic or another. This review focuses primarily on UV spectroscopic signatures of this variability and on studies that have exploited it to reveal the presence of large-scale velocity and density structures in the winds. The review includes demonstrations that the physical structures responsible for the variations appear to be large, spatially coherent and rooted in the stellar photospheres. Furthermore, evidence is summarised that these structures can be optically thick to resonance line radiation and are dense enough to modulate the X-ray fluxes that originate in the wind. Also discussed, is how the structures can affect wind line formation and, as a result, the precision by which mass-loss rates can be determined. The potential of millimetre and radio continuum observations to constrain the radial stratification of wind clumping is highlighted, as a future path to reliable empirical determinations of massive star mass-loss rates.

**Keywords:** Massive stars; mass-loss; stellar winds; clumping; UV spectroscopy

## 1. Introduction

Massive stars drive powerful stellar winds that inject substantial mechanical and radiative energy into the interstellar medium. This feedback directly shapes how massive star clusters and their host galaxies evolve. Additionally, wind-driven mass loss is a key factor in the evolution of single massive stars, and thus the nature of the neutron star or black hole end-state. The determination of reliable mass-loss rates, which in turn is founded on an understanding of the observational diagnostics, is thus critical across may areas of massive star astrophysics. The stellar winds are driven by radiative pressure on metal lines ([1]). Because the optical spectra of O stars are dominated by photospheric lines or recombination lines (which only sample the very base of the stellar wind in OB stars), empirical stellar wind research has concentrated substantially on UV resonance lines, which are formed throughout the wind. This review therefore also primarily focusses on diagnostics and perspectives derived from UV spectral lines.

Starting in about 2000, it became apparent that the values of observationally derived mass-loss rates founded on different spectral diagnostics were discordant by factors of 10 or more (e.g. [2], [3], [4], [5], [6], [7]), with most differing strongly from theoretical expectations (e.g., [8]). The heart of the problem is that OB star winds are far more complex than the homogeneous, spherically symmetric flows envisioned by [1]. Instead, there is overwhelming evidence that they contain structures which can be quite small or very large. The two observational results that can only be explained by structure are the variability of UV wind lines and the presence of X-rays. The UV absorption spectral line variability, in particular, is essentially universal. Figure 1 demonstrates the substantial temporal changes that may be evident across a range of OB stars.

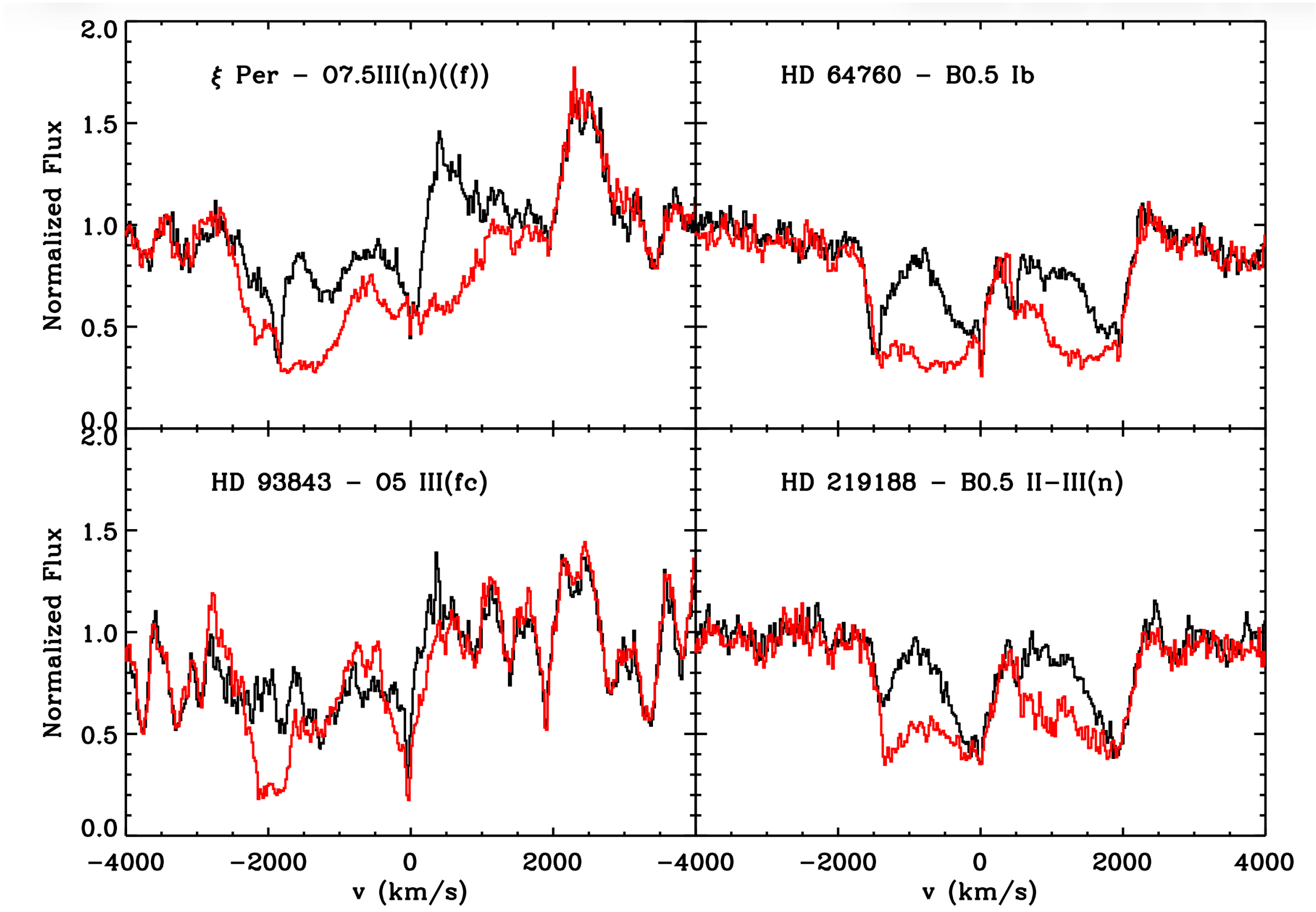

**Figure 1.**

Examples of large variability in the International Ultraviolet Explorer (IUE) Si IV λλ1394,1403 resonance line absorption profiles in four OB stars with a range in spectral types. In each case the black and red line profiles demonstrate changes occurring several days to a few weeks apart.

The wind-formed UV line profiles displayed in Fig. 1 are typical examples for OB stars. However, the morphology of the Doppler-shifted absorption troughs seen here is clearly not that of 'textbook' P-Cygni profiles; i.e. the absorption troughs are not smooth, but show optical depth features. Specifically, localised narrow absorption components (NACs), are commonly seen in single (snapshot) UV spectrograms (e.g. [9], [10]). The NACs occur at the blue edge of the absorption trough, near the wind terminal velocity. While their strengths may change with time, the NACs are mostly relatively fixed in velocity. Subsequent time-series IUE observations revealed that the NACs are spectroscopic end-products of temporal features, known as discrete absorption component (DACs). The DACs arise from much lower velocities as broad absorption features, which accelerate into narrower features at higher velocities, asymptotically approaching the terminal wind speed ($v_\infty$), typically ~2000 km s$^{-1}$. An example of this progression from DACs to NACs is show in Figure 2, for the O7.5 III((f)) star, 68 Cygni from [11]; see also Sect. 2.

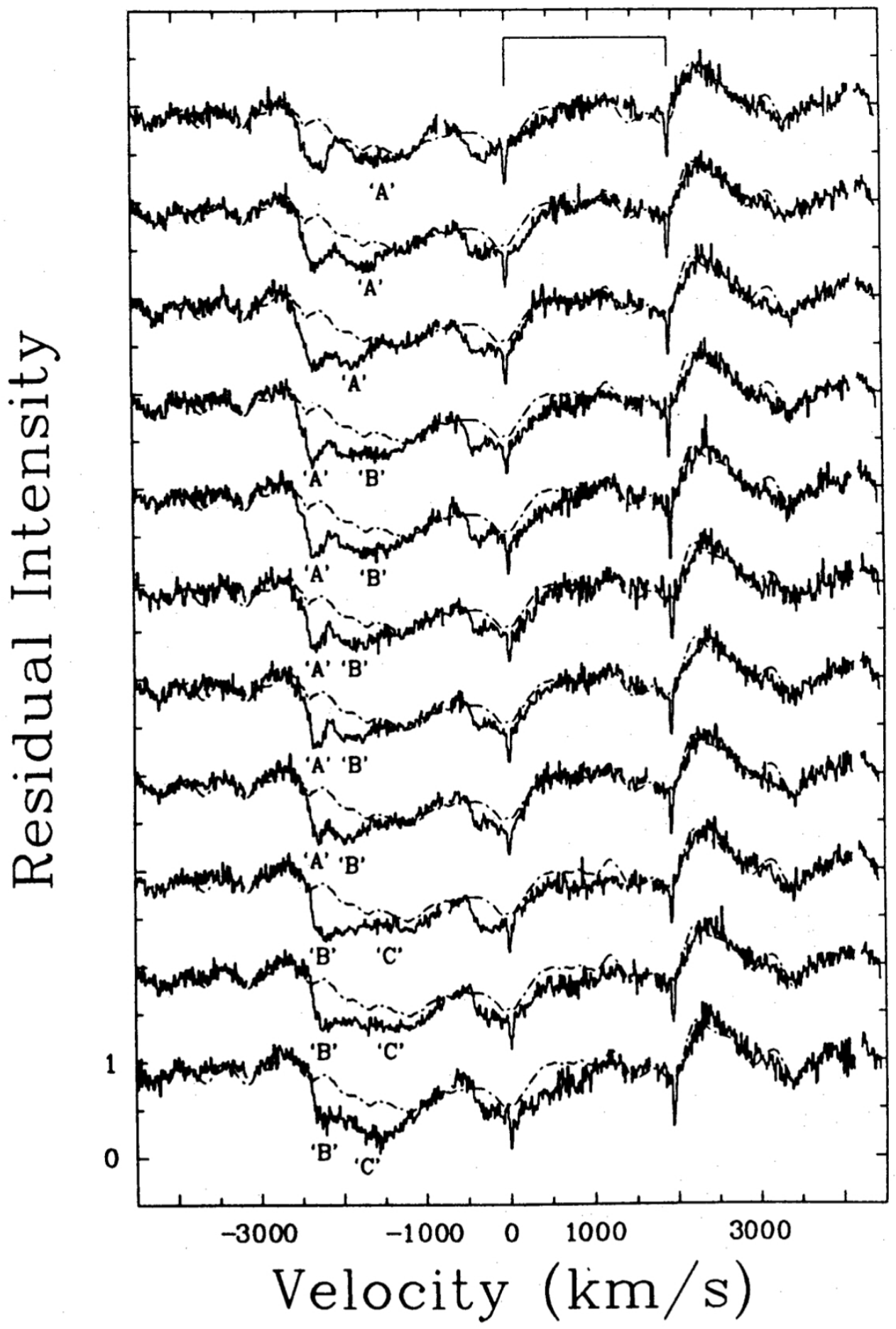

**Figure 2.**

Example from [11] of the progression of DACs to NACs in the IUE Si IV λλ1394,1403 absorption profiles of 68 Cygni. Three separate episodes are identified over about 80 hours. The bar at the top marks the Si iv doublet separation. For reference, a constant model P-Cygni profile is also shown (dot-dash line). Credit: Prinja, R.K. & Howarth, I.D., 1998, MNRAS, 233, 123, reproduced by permission of OUP and RAS.

Until we unravel the details of these flows, we cannot be confident in our use of observational diagnostics to infer mass-loss rates. Consequently, it is imperative to understand the origin and nature of density and velocity structure of the winds; importantly, time series of wind lines provide a unique and powerful method to probe these structures. In particular, determining whether the variability seen in the time series is due to material originating near or at the surface of the star and then propagating outward, or low speed material at large distances from the star which simply appears at low velocity, has been a key goal of several studies discussed below. Should the origin of all of the features be tied to the stellar surface, this would have implications about the structure of the winds (e.g. how does the component of the wind which contains small-scale clumps, and shapes saturated wind lines, co-exist with large scale structure), the structure of the photosphere (e.g. what causes the photospheric irregularities that give rise to the different wind flows), X-ray production (e.g do X-rays originate mainly in small, spatially incoherent shocks or along the interfaces of large-scale structures). These implications impact theoretical derivations and simulations based on a smooth wind.

## 2. Wind line variability

Large structures manifest themselves in time-series data (primarily collected by the International Ultraviolet Explorer, IUE, satellite mission) of unsaturated UV wind lines as discrete absorption components (DACs), which indicate large, coherent structures propagating through the winds (e.g.,[11], [12], [13]; see Fig. 2). Similar features are observed in LMC and SMC O stars ([14], [15]) and in central stars of planetary nebulae (CSPNe) ([16]), thus suggesting they are a universal property of radiatively driven flows.

The empirical diagnostic and analysis of large-scale wind structure is relatively simple, and is commonly based primarily on inspection of dynamic spectra. Dynamic spectra are images of time ordered spectra normalised by the mean spectrum or minimum absorption spectrum of the time-series. Figure 3 shows dynamic spectra of ζ Pup (O4 If(n)) and ξ Per (O7.5 III). The dynamic spectra are powerful tools for assessing the presence of features moving through spectral lines in both time and wavelength (or velocity). Notice that the variability is almost entirely confined to the absorption portion of the spectra. This is because the wind absorption is affected only by the wind material in a cylinder of radius $R_\star$ between the observer and the stellar disk, making it sensitive to localised variations in the structure of the wind at each velocity, $v$. Note that if the velocity of the wind flow is not a monotonic function of radial distance, r, the mapping between $v$ and r may not be one-to-one. Three features in Figure 3 are particularly important: 1. the range in absorption change is large, exceeding 50%; 2. there are rapidly evolving, DACs that repeat at regular intervals; 3. there is variation beyond the terminal velocity, $v_\infty$ = 1500 km s$^{-1}$.

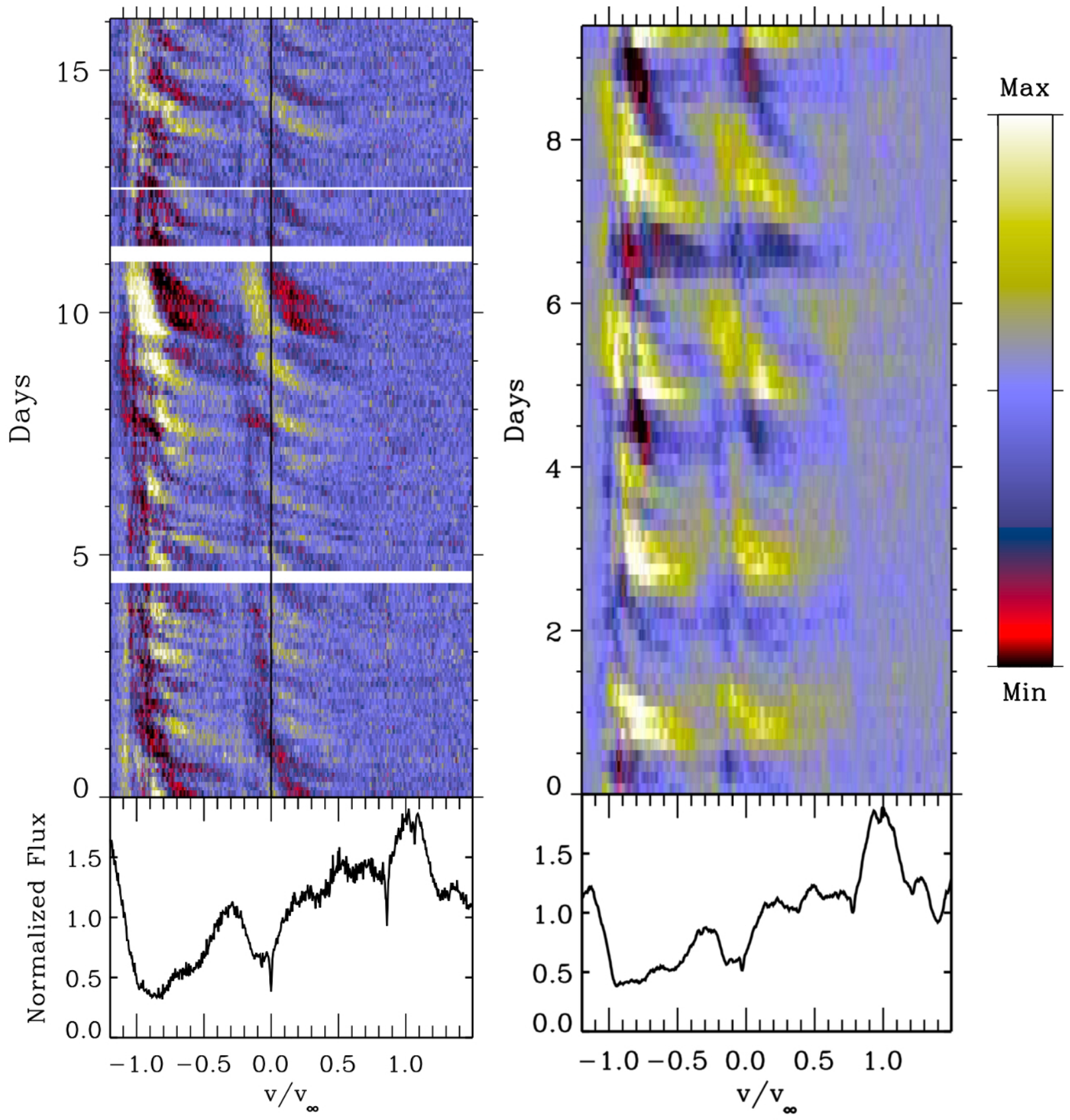

**Figure 3.**

Dynamic spectra of the Si IV λλ1394,1403 Å doublet resonance line, showing the progression of multiple DACs in ζ Pup, O4 If(n) ; left-hand panel) and ξ Per (O7.5 III; right-hand panel). The velocity is normalised by the terminal velocity ($v_\infty$) and is relative to the blue component of the doublet. The spectra are normalised by a minimum absorption (maximum flux) spectrum so that changes appear as absorption. (The quotient flux is show in the vertical bar on the right, which ranges from 0.4 to 1.2.)

More broadly for OB stars, the properties of coherent structure seen in the full comparison between different lines, (i) permits tests of whether every absorption feature observed at low or near-zero velocity connects to a feature at high velocity, (ii) reveals "banana" shaped patterns, which are the accepted signatures of structures tied to the stellar surface, (iii) potentially establishes connections between the wind features and photospheric line changes, (iv) constrains any instances of features formed (at intermediate velocity) in the wind: if all features can be traced back to the surface and they move through the profile monotonically with velocity, this would imply that we only see evidence for porosity in the dynamic spectra. (As opposed to structure that is in velocity space and arises from a non-monotonic velocity law; i.e. 'vorocity'), and (v) tests the notion that features in dynamic spectra must be huge, with large lateral extent in order to remain in the line of sight for days, persisting to very large velocity.

### *2.1. Rotational modulation*

IUE satellite spectra have revealed systematic behaviour of wind variability. The most well studied objects exhibit cyclical variation in the stellar winds which has a causal connection to the photosphere and where the time-scales of the wind changes are commensurate with the rotation periods of the OB stars (e.g. [12], [17], [18], [19], [20]). The physical interpretation of this rotational modulation is the action of spiral-shaped structure in the wind density distribution, caused by perturbations at the base of the wind and carried by stellar rotation (see Sect. 3).

All OB stars may have localised surface structure, such as spots caused by, for example, magnetic fields or non-radial pulsations (NRPs). However, there is probably some minimum combination of wind strength, rotation rate and viewing angle needed for the difference in conditions at the base of the wind to create a stable pattern in the wind and for an observer to see it (e.g. [21]). For example, there are numerous IUE spectra of 10 Lac (O9 V; it was used as an IUE standard), and its C IV profile is strong enough that you'd expect to see some variability, but don't. Though there is some blue-shifted absorption in N V ([12]). However, 10 Lac has a projected rotation velocity, v sin(i) ~ 15 km/s. In contrast, some B supergiants have low v sin(i)s, but do show DACs. They have strong winds and sin(i) could be large since their rotational velocities must be low. Inclination, sin(i), plays a role because the DAC activity is probably strongest near the equator (where the transverse velocity is largest and, possibly, spot-related amplitudes are biggest) and if sin(i) is small enough, then DAC activity might be present, but you just don't see it if your line of sight to the star is near the pole (the same argument applies to NACs).

### *2.2. Variability at very low velocity*

Modulations in the wind line intensity at low velocity provide important information about the origin of spiral structures. In specific cases where this region of strategically selected UV lines samples the wind close to the star, it reveals information about the footprint of the structure. There are two key spectral lines in the UV that can potentially diagnose very low velocity, near-surface, wind variability:

*(i) Excited state lines* – Determining the radial position of an absorption feature in the UV wind-formed line from its velocity alone is generally not straightforward. For instance, a low-velocity feature might originate from material accelerating near the stellar surface, or from distant, slow-moving wind material that is re-accelerating due to shocks. While this uncertainty always affects resonance lines, that is not the case for lines arising from an excited state. An excited state wind line arises from an allowed transition whose lower level is the upper level of a resonance transition (typically below 900 Å). They frequently appear as wind lines in O stars with strong mass fluxes. A commonly observed UV excited state lines is N IV $\lambda$1718, whose lower level is the upper level of the N IV $\lambda$765 resonance line. Because a strong EUV radiation field is required to populate an excited state line, these lines can only exist close to the star (this is what gives excited state lines their distinctive shapes). Since the population of the lower level relies on a strong radiation field, the absorption profiles of these lines are strongest at low velocities, near the photosphere, and quickly weaken at higher velocities (see, [22]). Instances where an absorption feature clearly migrates from low velocity in an excited state line to high velocity in a resonance line, indicate that the perturbation causing the optical depth change originated near the stellar surface and extended outward into the wind.

*(ii) Triplets* - Two well known, strong UV triplets that occur in OB stars are the C III 1175 Å and the Si III 1300 Å triplets. Triplets are interesting because their lower levels are populated by collisions and recombinations, which require a dense environment. Consequently, wind effects are only observed in stars with massive winds. The C III and Si III triplets become shifted, asymmetric and C III can even develop a P-Cygni profile. In dense winds, any evident variability of the triplets would indicate the presence of inhomogeneity, near the base of the wind.

IUE time-series data of 5 O stars were presented in [23], where they spliced together dynamic spectra of the excited state line N IV $\lambda$1718 at low velocity with those of Si IV $\lambda\lambda$1394, 1403 at high velocity in order to examine the extended, temporal evolution of wind line features. An example from their study is shown in Fig. 4. Every absorption feature observed at low velocity connects to a feature at high velocity; they advect, uninterrupted, toward the wind terminal velocity ($v_\infty$). Some of the spliced Si IV–N IV dynamic spectra reveal banana-like shapes, which are the accepted signatures of co-rotating interaction regions (CIRs) tied to the stellar surface (see Sect. 3). Ordinarily, features reaching to v $\simeq$0 does not necessarily imply they originate at small radii. However, as outlined above, by showing the features are present in excited state lines, which require an intense radiation field, implies that they must originate close to the star.

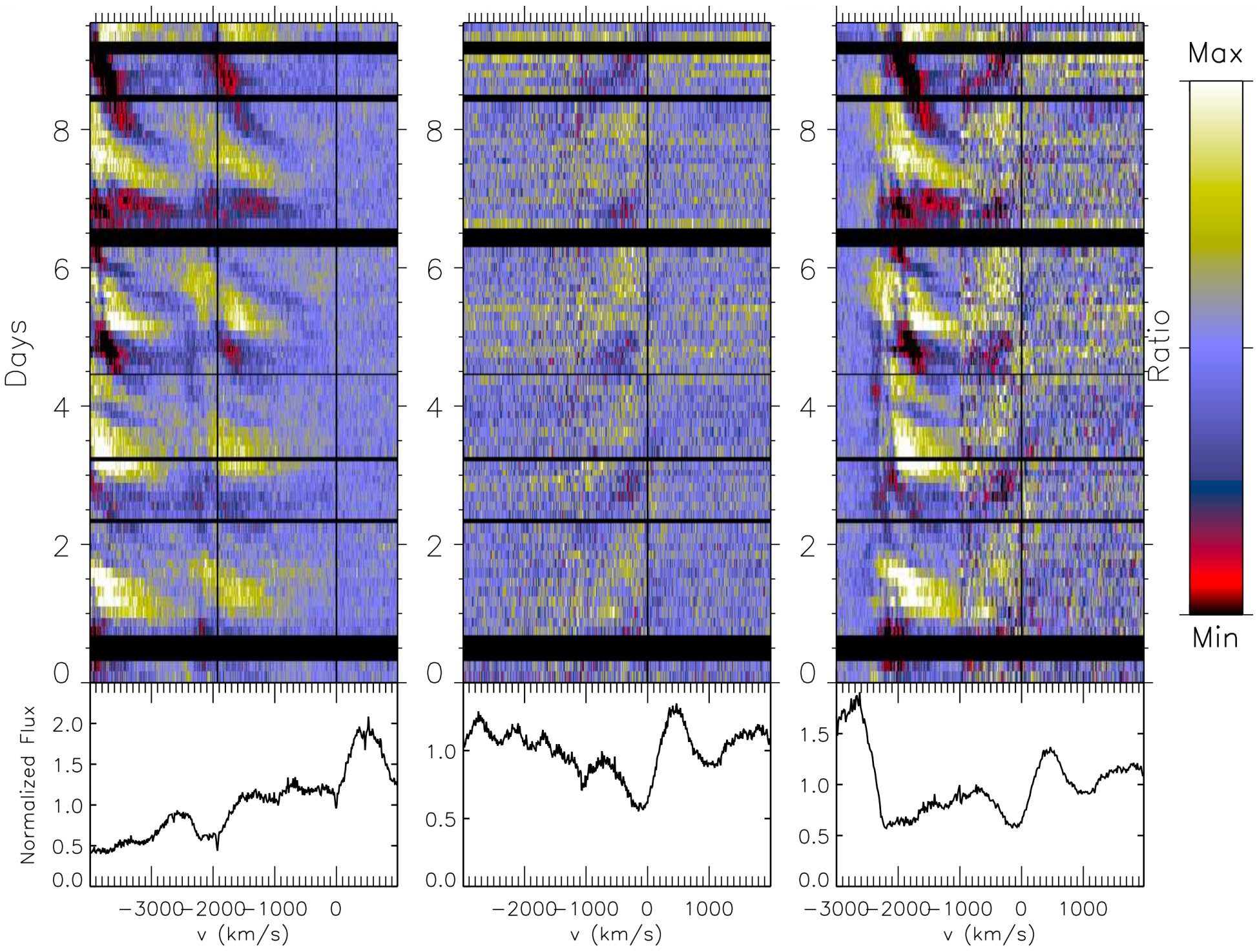

**Figure 4.**

Dynamic spectra (in velocity, $v$) of ξ Per for Si IV $\lambda\lambda$1394,1403 Å (left), N IV 1718 Å (center), and the spliced spectrum of Si IV–N IV (right) [23]. Note, the dynamic spectrum on the left (Si IV) is relative to the rest velocity of the *red component*, Si IV 1403 Å. Mean spectra used to normalise the images are shown below each panel. Credit: Massa, D. & Prinja, R.K. 2015, ApJ, 809, 12, reproduced by permission of the AAS.

Figure 5 shows dynamic spectra of three lines in HD 91969 (B0 Ia). It demonstrates how variations in the almost totally saturated Si IV doublet (left panel) join those in the weaker, intermediate velocity Al III 1855,1863 Å doublet (middle panel) which, in turn, appear to be tied to variations in the strong Si III $\lambda$1300 triplets at effectively v= 0. This is particularly important since the triplets must originate in a dense environment. Consequently, there is little doubt that the features originate in a dense region, or effectively the photosphere of the star.

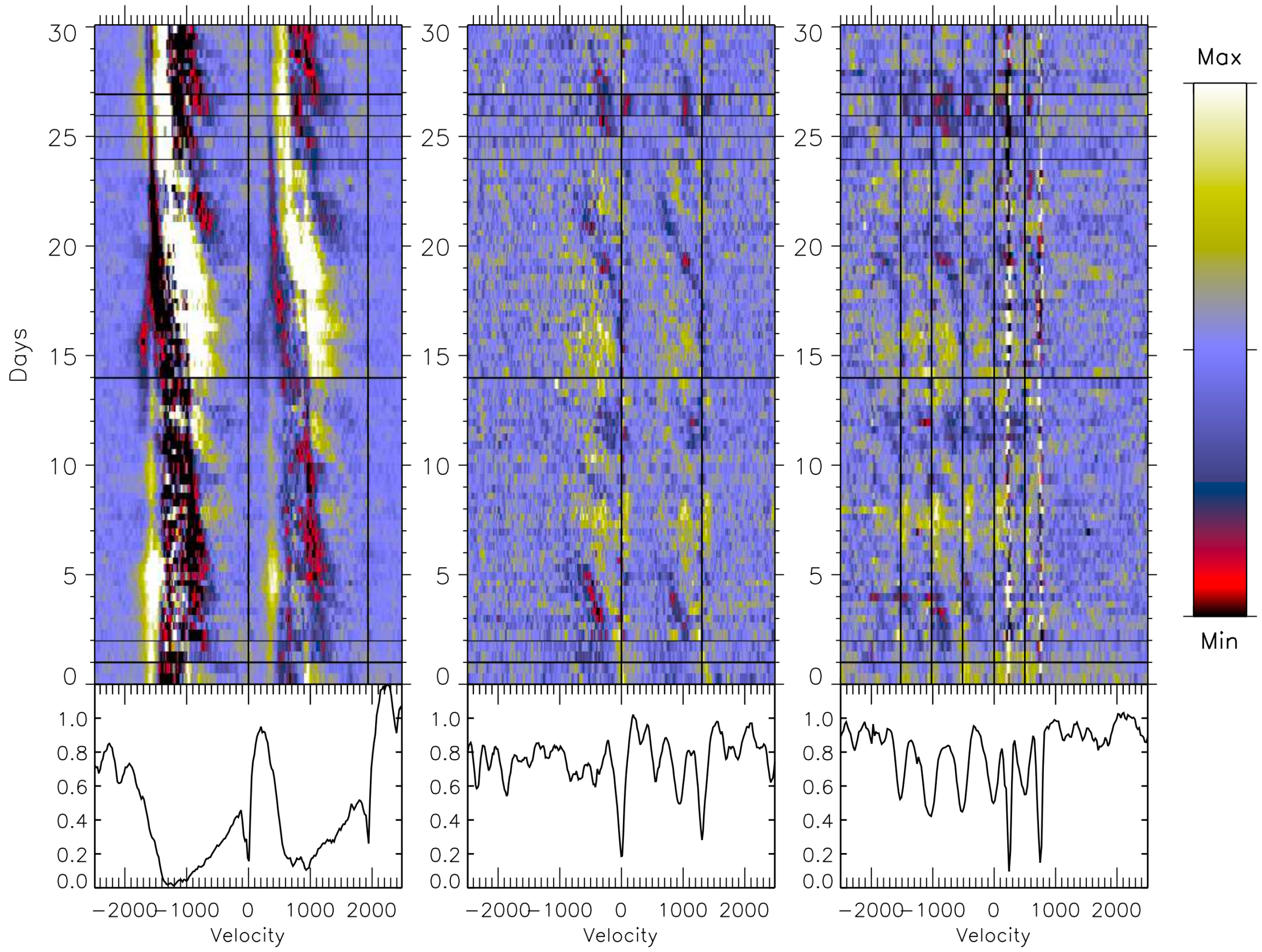

**Figure 5.**

Dynamic spectra of the Si IV $\lambda\lambda$1400 Å and Al III $\lambda\lambda$1860 Å resonance doublets and the Si III $\lambda\lambda$1300 triplets in HD 91969 (B0 Ia). There are 82 spectra with a mean sampling of 8.7 hr. This figure shows a very important property: The variable features in a strong line are linked to features in the weaker lines which, it turn, extend to zero velocity, and align with the triplets.

### 2.3. *Super terminal velocity absorption*

Super terminal velocity absorption is the absorption that occurs at velocities *larger* than $v_\infty$. Such absorption is evident in both Figs. 3, 4 and 5. For a classical, smooth wind, such absorption should not exit. Consequently, it has often been taken as evidence for the Line-deshadowing instability (LDI, [24], [25]), whereby parcels of the wind may be accelerated to velocities beyond $v_\infty$, before shocking and creating X-rays. However, most LDI models (e.g. [26]) predict that the LDI fragments should have relatively small geometric cross sections. This means that the signature of such structures should appear instantaneously and not last very long before either relaxing or moving out of the line of sight between the observer and the stellar disk. However, for the dynamic spectra examples in Figs. 3, 4 and 5 (see also [12]), the super-$v_\infty$ absorption appears to be temporally coherent over several days, and in some cases modulated on similar time-scales to the DACs. The P-Cygni 'blue edge' variability is undoubtedly not straightforward, and more systematic UV studies are needed to address this important issue.

*2.4. A signature of binarity?*

The UV time-series contain additional temporal features, that are currently enigmatic. A key example are very slowly evolving absorption components that co-exist with more rapidly migrating features. An instance can be seen in Figure 6 for HD 64760 (B0.5 Ib): A narrow feature is seen in Si IV migrating between -1000 to -1300 km s $^{-1}$ *over more than 10 days* and then possibly repeating. This very slow feature has no connection to the more horizontal 'periodic absorption modulations' [18] seen in this dynamic spectrum (that repeat over ~ 2 to 4 days, and are commensurate with rotation timescales.). A similar very slowly moving feature was noted by [13] for HD 164402 (B0 Ib; see their Fig. 3). When the data time span is extensive enough, these slow features are seen to evolve on time scales longer than the rotation period of the star and often first appear at *intermediate* velocity (unlike DACs). The origin of these much slower moving features is unknown. One interesting possibility may be that they are due to low mass companions embedded in the wind. Commonly, the OB stars reviewed here are not strong X-ray sources (see Sect. 5), such as high-mass X-ray binaries, with accreting compact objects interacting in the primary's stellar wind. Theoretical work suggests that many massive stars may be the products of binary mergers (e.g. [27], [28]). Examples of such companions have been discovered in Be stars by [29]. It is very intriguing to consider that the very slowly evolving feature seen in some OB stars (alongside the DACs) are in fact traces (i.e. wakes in the wind) of similar (sdO-B) stars in the UV spectra of the more massive stars. Further, multi-waveband spectroscopy and photometry would be worthwhile to explore this possibility.

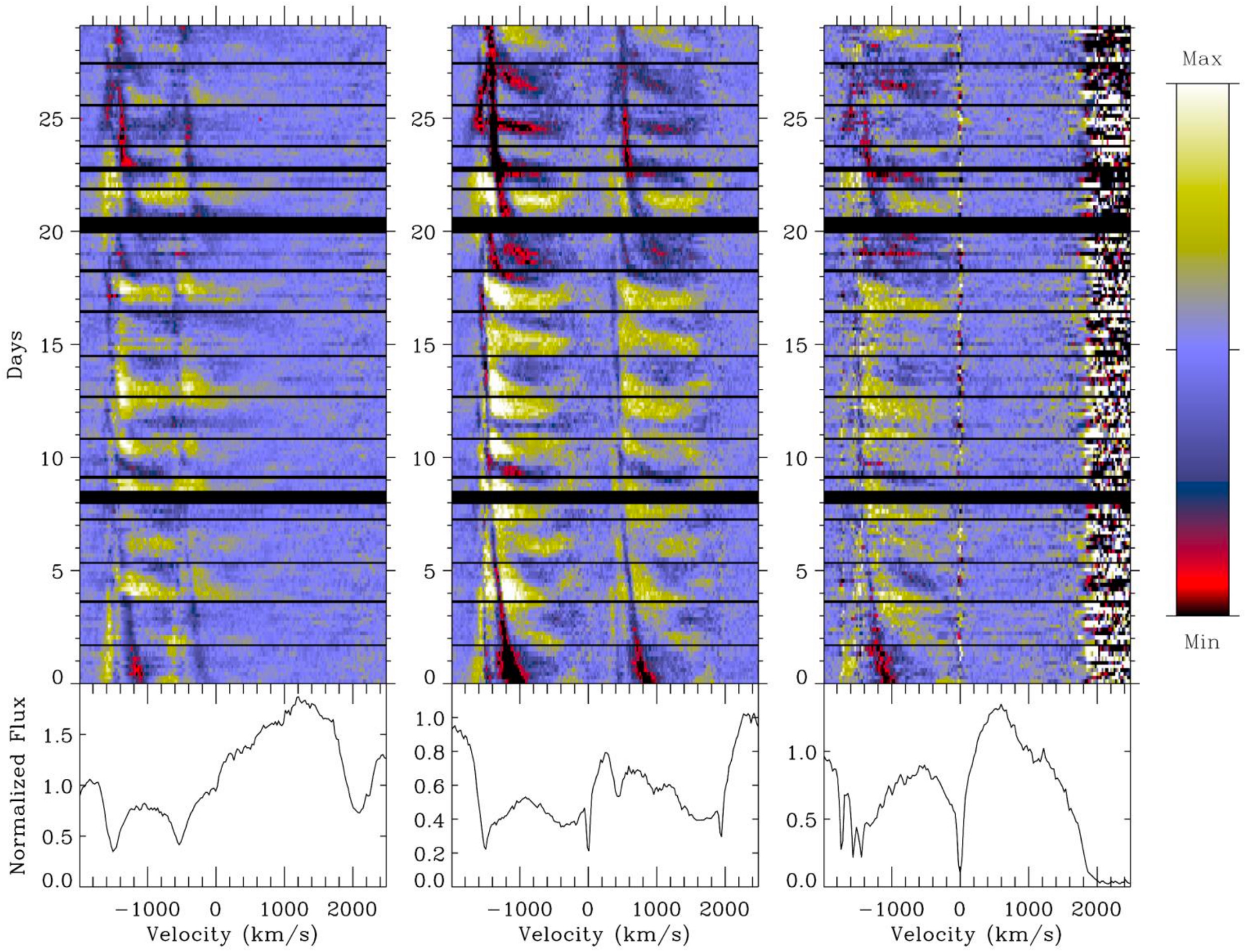

**Figure 6.**

From Left to right, N V λλ1239,1243 Å , Si IV λλ1394,1403 Å and Si III λ1207 Å dynamic spectrum of B supergiant HD 64760. Velocity is relative to the blue component for the doublets. Note the narrow feature seen moving between -1000 to -1300 km s $^{-1}$ *over more than 10 days* and then possibly repeating.

## 3. Co-rotating interaction regions (CIRs)

While the absorption features due to DACs observed in UV time series propagate monotonically to higher velocity, they appear to move more slowly than the velocity law of the wind (determined from the shapes of saturated wind lines) would predict. [30] presented a plausible account for these features, suggesting that they could be explained as the signatures of co-rotating interaction regions (CIRs) within the stellar winds. This phenomenon is similar to structures observed in the solar wind, where different regions of the Sun's surface generate low- and high-speed solar wind streams that accelerate at varying rates. As the star rotates, these differing sectorial flows collide, creating a spiral-patterned interface and a corresponding plateau in velocity. Consequently, as the CIR passes through the line of sight between the observer and the stellar disk, it appears as an absorption feature moving slowly outward through the wind. This explanation has considerable appeal, since it would not only explain the slow acceleration of the features, but also their large spatial coherence, implied by the persistence of the absorption features along the line of sight, often for several days. Figure 7 is a schematic of how the signature of a spiral pattern originates.

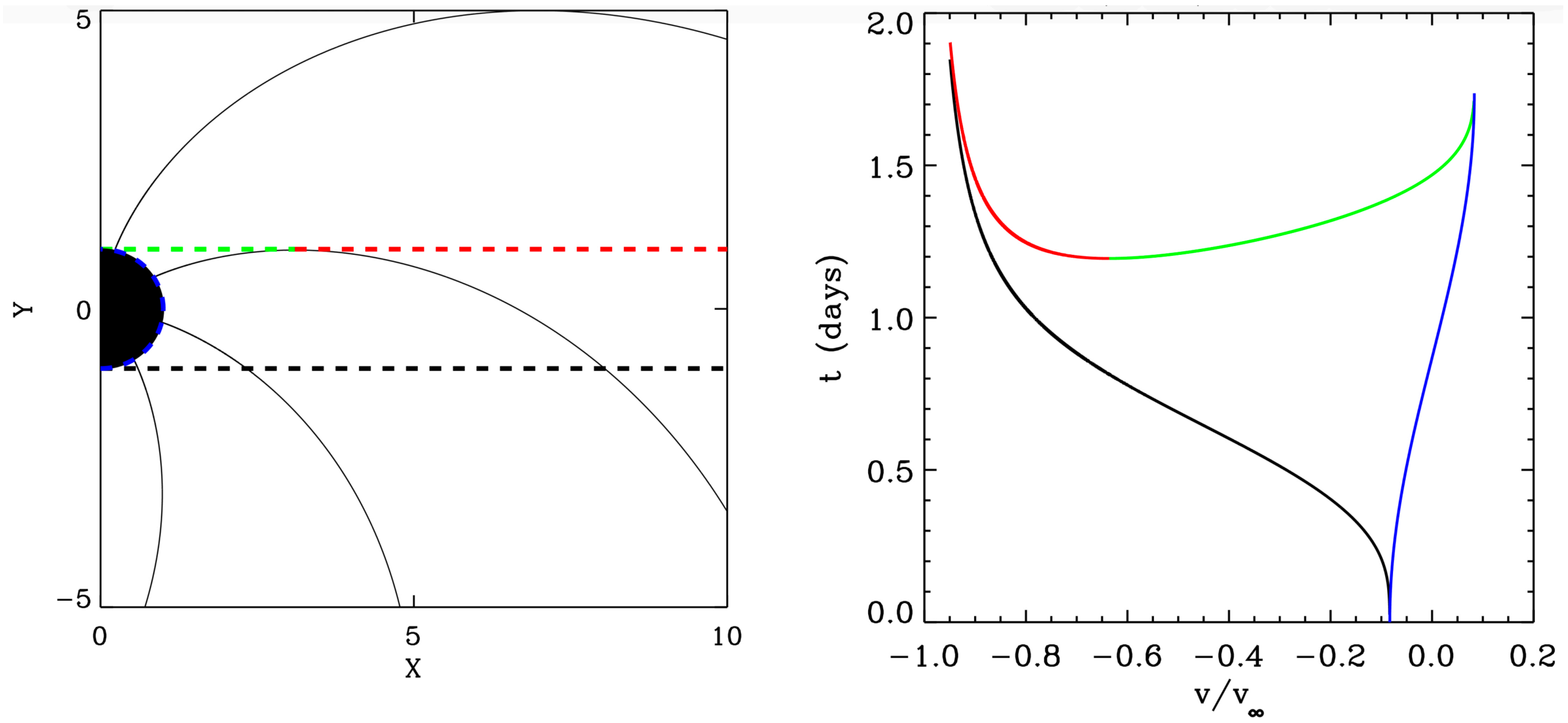

**Figure 7.**

Left: A spiral pattern shown at different times; the X, Y units are stellar radii. The black dashed line at Y = −1 is where the pattern first enters the line of sight. The red dashed line at Y = 1 denotes where it exits. The spiral exits first at intermediate velocity. The final boundary is the surface of the star, denoted as a blue dashed curve. Right: The region in a dynamic spectrum where a spiral can absorb. The boundary colours correspond to the spatial ones defined at left. The parameters used to create this plot correspond to those of ξ Per.

[31] showed how the spiral structures produced by co-rotating interaction regions, can explain the rapidly evolving DACs by introducing bright spots near the equator. The CIR model for rapidly evolving DAC formation was applied to HD 64760, B0.5 Ib ([18], [23]) and ξ Per, O7.5 III(n)((f)) ([19], [32]), and its signature appears in the dynamic spectra of most long duration time series (e.g., [13], [23]). The CIR model predicts spiral structures consisting of density

enhancements of ~2 together with velocity plateaus which can increase the Sobolev optical depth ($\tau_{rad}$) by 10 to 100 (e.g. [21]). The passage of CIR spiral arms across the line of sight to the stellar disk accounts for the wind line variability. While CIRs may have little effect on the mass-loss rate, they can strongly affect the observational diagnostics used to determine it. Although the spiral structures produced by CIRs explain well the rapidly evolving DACs, there are problems with this model. First of all, the spiral structures appear to extend to the photosphere ([23]), while the CIR spirals take time to form, usually not developing until intermediate wind velocities. However, other sources of spiral patterns have been suggested which might be able to create deeply rooted spiral patterns, including magnetic spots or non-radial pulsations (e.g. [33], [34]), but these models have not been fully developed, and it seems that magnetic dipole fields are ruled out as a source in most OB star cases ([35]).

## 4. Translating the observed variations into changes in physical parameters

Some sort of model is required to translate the observed variations into changes in physical parameters. Generally, detailed studies of individual UV and optical spectra of OB stars have required the development of spherical, non-LTE model atmosphere codes incorporating the effects of metal lines, namely CMFGEN ([36]), WM-BASIC ([37]; UV studies only), PoWR ([38]), and FASTWIND ([39]). However, whilst extensively and successfully deployed for deriving fundamental stellar parameters, these models do not incorporate the localised velocity and optical depth structure shown in this review to be observed in the wind-formed UV lines, such as DACs and NAC. Essentially, repeated calculation of any CMFGEN, PoWR etc. model cannot match the absorption shape or variability due to migrating DACs seen in repeated UV observations of OB stars. (Nor, indeed, the narrow absorption components, NACs, that are ubiquitously evident at the terminal velocity in unsaturated P-Cygni profiles.)

The simplest, and least restrictive alternative method is the Sobolev with extra integration (SEI) model, originally developed in its standard form by [40]. This model can be viewed as an inversion of the profile to obtain the radial optical depth for a line, $\tau_{rad}(w)$, where $w = v/v_\infty$. It assumes spherical symmetry and a monotonically expanding wind. The SEI model is based on a "core-halo" representation of the star and its wind. This assumes that the photosphere of a "bare" star is embedded in a rapidly expanding wind. The model works very well for stars without enormous mass-loss rates and is particularly accurate further out in the wind, and accounts for the interaction of closely spaced doublets. While SEI is nether self-consistent or exact, it is flexible and computationally swift, allowing it to be ideal for use in non-linear least squares routine. A successfully applied modification of the standard SEI method has been to model $\tau_{rad}(w)$ in a set of 21 independent velocity bins whose magnitudes were adjusted to obtain the best fit (e.g. [2], [41]). This method has the advantage of avoiding biases due to preconceived notions about the functional form of the distribution of $\tau_{rad}(w)$. It also simplifies the fitting procedure, since $\tau_{rad}(w)$ is the only free parameter once the velocity law and photospheric spectrum are fixed. The radial optical depth of the wind, $\tau_{rad}(w)$, gives the run of optical depth as a function of radius. It is directly related to the physical parameters that determine the flow and can be parameterised as,

$$\tau_{rad}(w) = \mathrm{Const}\frac{\dot{M}}{R_\star v_\infty^2}\frac{q_{i\,(w)}}{x^2 w}\left(\frac{dw}{dx}\right)^{-1}\lambda f$$

where $q_{i(w)}$ is the fraction of an element in ionization state i at velocity, $w$, $\dot{M}$ is the mass-loss rate of the star, $f$ is the oscillator strength (see below) and Const is a numerical coefficient that contains other atomic parameters.

### *4.1. Diagnosing optically thick, large-scale structure*

In addition to the normal parameters (e.g. velocity law, input photospheric spectrum, optical depth bins as above), an interesting technique for UV doublets is to allow the ratio of the oscillator strengths of the blue and red components of the doublet, $f_B/f_R$ (termed f–ratio), to be a free parameter. For a source obscured by a *uniform distribution* of optically thin plasma, the ratio of optical depths determined from the observed absorption lines would equal the ratio of the oscillator strengths of the two doublet components. The idea of using f–ratios as markers for a clumpy medium that partially covers a source is relatively simple. A medium that is optically thick and only partially covers the source will produce a set of doublets with nearly equal strength, but only slightly different equivalent widths, and not very deep. As a result, they resemble absorption by a doublet whose f–ratio is nearly unity, instead of $f_B/f_R$. [42] showed that, for most stars, the best fit to a doublet wind profile is obtained by allowing the f–ratio to vary. They provide evidence for apparent f–ratios closer to 1, instead of the expected value of 2, yield better fits for Si IV wind lines. This indicates that the stellar surface is partially covered by optically thick and partly by optically thin structures. The profile is thus a combination of regions that would make the profile completely black in both components and regions where there is little or no absorption. The result is a profile that appears unsaturated (because its flux does not go to zero), but whose doublet components have nearly equal strength (this effect has been observed in Active Galactic Nuclei (AGN) absorption lines for many years, (e.g. [43]). The variable f–ratio technique can also be applied to repeated observations or to the individual spectra in a time series. Examples of variations in f-ratio derived from IUE time-series spectra of 4 OB stars are shown in Figure 8.

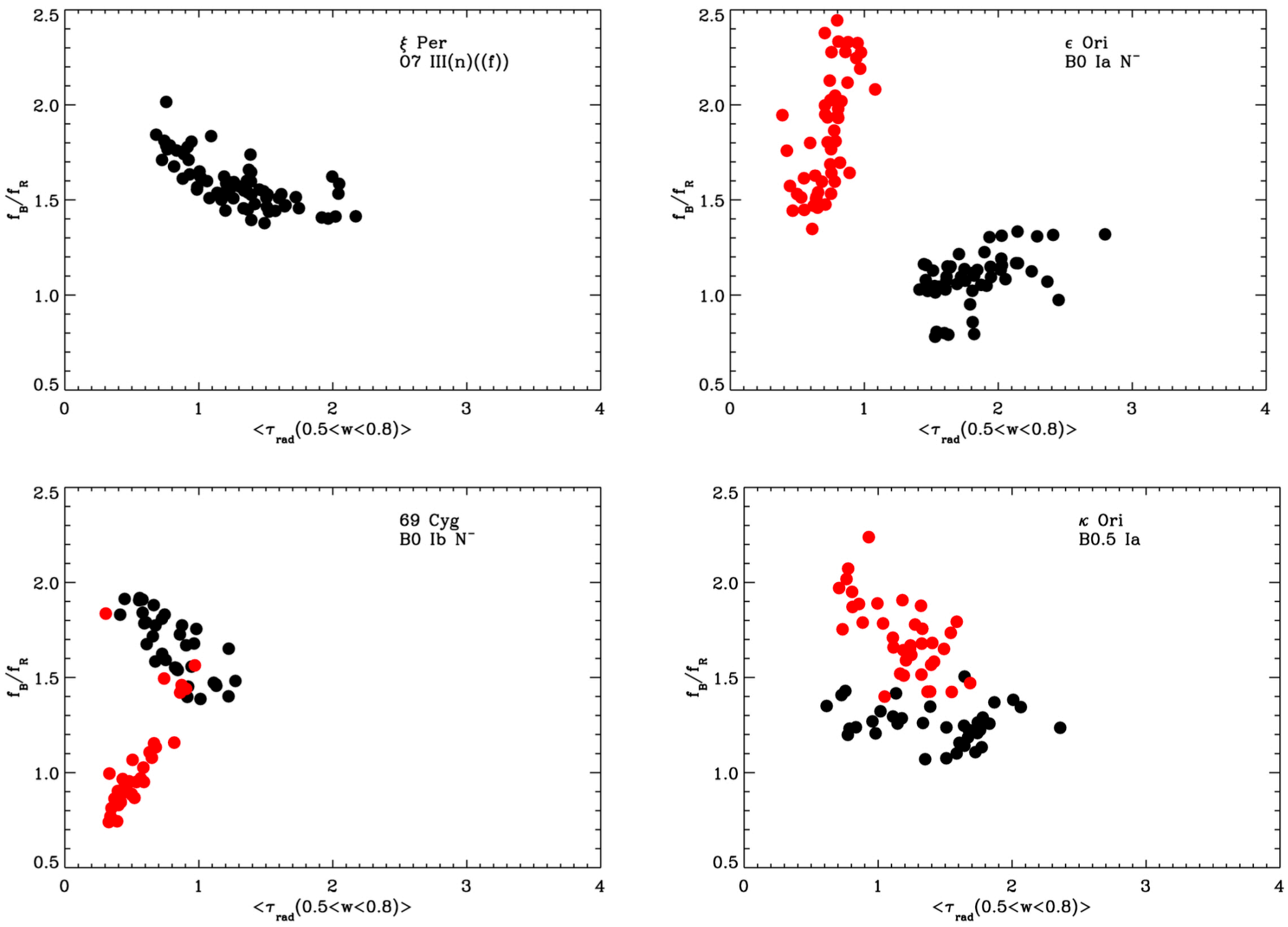

**Figure 8.**

Plots of f-ratios for Si IV (black) and N V (red, when well-developed, but unsaturated) versus the variations in $\tau_{rad}$ averaged over $0.5 \leq v/v_\infty \leq 0.8$, for the stars listed in each panel.

The results tell us how much of the variability is due to larger and smaller covering factors as opposed to larger and smaller ambient optical depths. Such results potentially provide valuable information about the geometry of the winds. Given that $\tau_{rad}$ spans all velocities and radii, the use of a single value for $f_B/f_R$ is a crude measurement of the covering factor. Nevertheless, it does provide a measurement of a poorly defined mean that can be a useful diagnostic. Broadly, these results indicate that the winds contain structures that are optically very thick and that the apparent absorption is actually a measure of how much of the stellar disk is covered by these structures at a given time. The doublet anomaly can be interpreted in terms of optically thick LDI fragments ([42], [44]). However, one must keep in mind that when optically thick structures are present, geometry matters. Therefore, we need to constrain the nature of the optically thick structures as much as possible. Until we unravel the details of these flows we cannot be confident in our use of observational diagnostics to infer mass-loss rates. Consequently, it is imperative to understand the density and velocity structure of the winds and time series of wind lines provide a unique and powerful method to probe these structures.

To truly understand the variability and its effect on the accuracy of parameters such as mass-loss rate ($\dot{M}$) derived *from a single spectral measurement*, we need to translate the observed flux variability into the variability of physical parameters. [44] performed an SEI analysis of repeated observations in order to quantify the intrinsic error in any mass-loss rate ($\dot{M}$) measurement derived from a UV wind line. While the individual derived optical depths may not be exact, their variance is shown to be reasonably robust. This study was primarily concerned about how fitting a profile obtained at one epoch can differ from fitting a profile at a different epoch. So in this case, for an assumed $v_\infty$, they allow all of the other parameters (including the f-ratios) to vary, in order to obtain the best possible fit to the profile. The distribution of $\dot{M}$ and its relation to variations in the other parameters can be examined. These variations quantify the intrinsic error inherent in any single mass-loss rate derived from a single UV observation. The derived rates have an overall 1$\sigma$ variation of about 22%, but this appears to differ with effective temperature, being as small at 8% for the hottest stars and up to 45% for the cooler stars. In addition, any single determination can differ from the mean by a factor of 2 or more. These results imply that mass-loss rates determined from non-simultaneous observations (such as UV and ground based data) need not agree, due to the intrinsic temporal variance of the spectroscopic diagnostics.

## 5. Modulated X-ray emission

The observation that all O stars emit X-rays also implies wind structure. X-rays cannot be produced radiatively in O stars, so their presence implies mechanical processes which, in turn, implies structure. It is widely accepted that the X-rays arise from the line de-shadowing instability, LDI (e.g., [45], [46], [47]). In LDI models, the wind fragments into randomly distributed structures which interact with each other and the ambient wind to produce X-rays throughout the wind. Some LDI predictions agree with observations. For example, X-ray emission line profiles (e.g., [48], [49],[50]) imply that the hot plasma moves with the same velocity as the cool wind, and that the X-rays are attenuated by the cool gas, as expected.

However, other LDI model predictions are at odds with observations. The model predicts strong short time scale stochastic X-ray variability ([51]) while none is observed ([52]). It also predicts that the strongest shocks (hottest plasma) should originate well out in wind, while observations indicate that it is very close to the photosphere ([53]). Further, there is growing evidence that X-ray fluxes vary on time scales of days ([52], [54], and [55]), which is contrary to LDI predictions. Could this modulation be caused by the spiral structures? If so, it would require far larger density enhancements than CIR models predict. Finally, it has been suggested that the LDI clumping could disrupt the large scale spiral structures needed to explain the UV time series ([21]), although [47] speculate that the LDI can produce both the spiral patterns seen in UV wind lines and the small scale structure responsible for the X-rays. Considerable work remains before this result is secure and specific predictions of its effect on mass loss - and the diagnostics used to determine it - can be quantified.

## 6. Millimetre and radio continuum constraints on clumping

While hot stars emit the bulk of their radiation in the UV and optical, at mm and radio wavelengths, thermal radiation from the stellar wind dominates; as a consequence, it is possible to employ mm/radio-continuum observations to estimate mass-loss rates ([56]). Powerful facilities such as ALMA, EVLA and e-MERLIN have delivered improved sensitivities, allowing three key physical advantages to be exploited: (i) the mass-loss determination is not dependent on the uncertain wind acceleration law, since at the radii at which the wind emits mm/radio-continuum radiation has already reached its terminal velocity, removing an important and ill constrained variable from the calculation; (ii) the free-free flux depends on wind density squared; (iii) the mass-loss estimates are not significantly affected by interstellar reddening.

The continuum emission at mm wavelengths is due to Bremsstrahlung (free-free) radiation and depends on wind density and so is sensitive to the degree of structure within the wind; i.e. a clumped wind produces more mm-continuum emission than a wind where the same amount of material is smoothly distributed. This is illustrated in Figure 9 (from [57]), which compares predictions for the emergent spectrum from the photosphere plus a smooth stellar wind for the canonical B supergiant ε Ori to observations from optical to radio wavelengths.

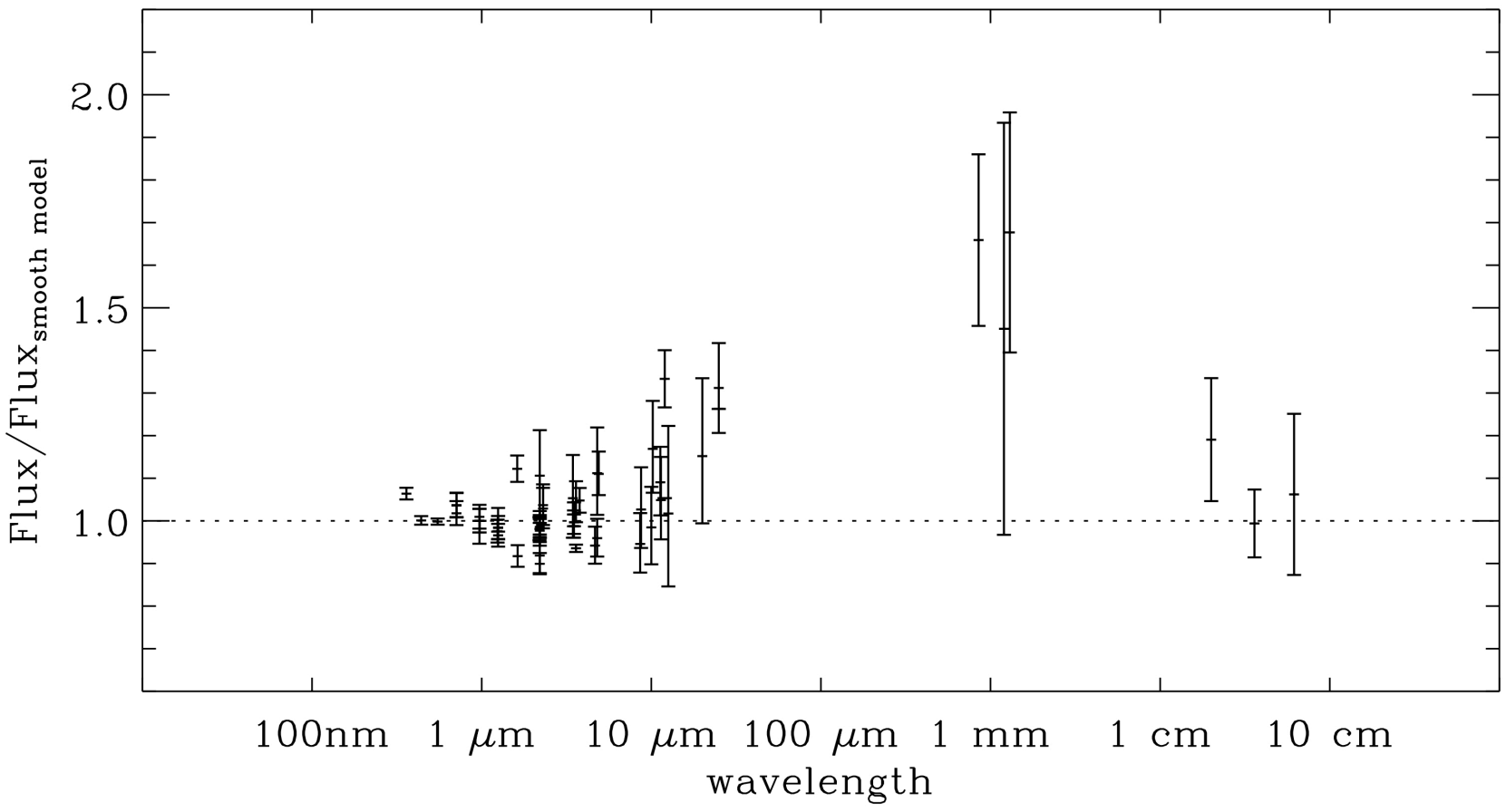

**Figure 9.**

Observed free-free (thermal) continuum fluxes in ε Ori (B0 Ia), normalised to a smooth wind model. Observations above the dotted line point to additional emission hat is not included in the smooth model. Credit: Blomme, R., Prinja, R., Runacres, M., & Colley, S., 2002, A&A, 382, 921 [57], reproduced by permission of EDP Sciences and ESO.

Furthermore, it is clear from mm observations that substantial sub-structure is present in the wind and if this is not accounted for - as in standard spectroscopic determinations of mass-loss rates - then a significant over-estimate will be made, which will propagate through into erroneous predictions for stellar evolution (i.e. the biasing of the production rate of neutron stars over black holes). In combination with spectroscopic observations, which constrain the composition, ionisation structure and terminal velocity of the wind, it is possible to employ mm-continuum observations to make a robust estimate of mass-loss rates, correcting for the effects of wind clumping.

An attempt to constrain the radial stratification of the clumping factor was performed by e.g. [6], [58] who analyzed H$\alpha$, IR and radio-fluxes, of a sample of early-type stars. A major result found was that in *weaker* winds the inner clumping factor ($r < 2\ R_\star$) has the same value as the outermost one, indicating similar clumping properties. In contrast, for *stronger* winds, the inner clumping factor is larger, by a factor of 3 - 6.

Radio and mm continuum observations of free-free thermal emission, at different frequencies, allow the outer wind characteristics to be radially scanned, thus providing perspectives on the clumping gradients in these regions. The effect of wind clumping in determination of mass-loss rates in empirically included via the clumping factor, $f_{cl}$. It is assumed that the clumping factor is uniform across clumps with a value given by $f_{cl} = \langle\rho^2\rangle/\langle\rho\rangle^2$, the inverse of the volume filling factor. (The angle brackets indicate an average over the volume in which the radio continuum at that wavelength is formed.) A smooth wind (i.e. no clumping) has $f_{cl} = 1$, whilst a value of $f_{cl} > 1$ indicates a clumped wind that for a given flux density reduces the mass-loss rate. Recent mm and radio studies have examined clumping gradients in the outer winds of WR, O and B supergiants in massive stellar clusters in the Mily Way Galaxy. For example, [59] (Westerlund 1); [60] (Arches cluster) and [61] (Arches cluster) have used ALMA and VLA measurements of individual stars to estimate clumping ratios at different frequencies (and line formation regions). The overall trend is a decrease in clumping with increasing radial distance.

## 7. Concluding remarks

There are several specific constraints provided by the extensive spectroscopic, and other, studied discussed in this review:

1. UV variations are due to organised structures sweeping across the line of sight to the stellar disk; as opposed to randomly placed clumps. The associated changes can be enormous in absorption line profile shapes from one observation to the next.
2. The characteristics of the variability implies a breakdown of spherical symmetry, radial homogeneity and possibly (but not necessarily) steady flow.
3. The DACs are temporally coherent and modulated on stellar rotational timescales, implying they are due to spiral structures rooted at the stellar surface. Eventually these structures pile up in velocity space, and are seen as NACs. Lack of activity in very low projected rotation velocity, v sin(i), stars could be due to small v or small sin(i).
4. The anomalous ratios of doublet UV resonance lines imply that at least some structures are optically thick and only partially cover the stellar disk at any time. This means that the mass-loss rates could be far larger than a straightforward interpretation of the profiles suggests.

5. There is detectable X-ray variability on the time scales similar to those evident in UV time-series. The implication is that the spiral wind structures may be dense, however precisely what the X-rays are measuring is dependent on how the they are distributed in the wind, which, in turn, is very model dependent.
6. The radial stratification of small-scale wind clumping can be constrained by scanning thermal wind continuum emission at mm and radio wavelengths. The trend is for a decreasing clumping gradient toward longer wavelengths, i.e. in the radio regime.

Together, these results present problems for inferring mass-loss rates from current models of P-Cygni profiles or other mass loss indicators, which are all subject to mis-interpretation if large density variations are present. Thus, to derive reliable mass-loss rates, we first need to understand the nature of the structures in stellar winds. Further modelling is essential. For example, we do not know how much mass is in the structures responsible for the DACs. Sobolev optical depth depends on both density and velocity gradient. So the wind structure may be minor enhancements with large optical depths due to velocity plateaus, or they are major density enhancements. The UV resonance line doublet ratio anomaly implies that only a portion of the star is covered at any one time. This means that the mass-loss rates could be far larger than a straightforward interpretation of the profiles suggests – when optically thick structures are present, *geometry matters*.

The development of new or more extensive models will advance our understanding of mass-loss and wind structure in massive stars. For example, the development of 3D model atmospheres for hot, massive stars with winds remove assumptions of spherical symmetry with a stationary outflow (e.g. [62]). Connecting these theoretical simulations with observations offers promising avenues for future work. For example, the ULLYSES/XShootU projects (e.g. [63]), which combine Legacy HST UV spectroscopy, optical and near-IR spectra, and line-synthesis modelling, are advancing our understanding of stellar winds and mass-loss in the lower metallicity environments of the SMC and LMC. Meanwhile, deeper examination of the archival UV data, of the broadest OB, WR and CSPN parameters can provide additional constraints on the nature of wind structure (both geometric and kinematic) that lead to the design of new tractable observational tests, and ultimately a self-consistent description of mass-loss rates in hot stars.

**Acknowledgments:** This review is dedicated to Derck Massa, who passed away in July 2026. Over several decades, Derck made significant contributions to advance our understanding of the stellar winds of OB stars. I thank Alex Fullerton for his very helpful comments on an earlier version of this manuscript, and two anonymous referees for providing further comments.

**Conflicts of Interest:** The author declares no conflicts of interest.

**Data Availability Statement:** The original data presented in the review are openly available [see DOI/URL].

## Abbreviations

The following abbreviations are used in this manuscript:

CIRs co-rotating interaction regions

CSPNe central stars of planetary nebulae

DACs discrete absorption components

IUE International Ultraviolet Explorer

LDI line-deshadowing instability

NACs narrow absorption components

NRPs Non-radial pulsations

SEI Sobolev with extra integration

WR Wolf-Rayet